# Growth-Etch Metal-Organic Chemical Vapor Deposition Approach of WS$_2$ Atomic-Layers

Assael Cohen,[†#] Avinash Patsha,[†#] Pranab K. Mohapatra,[†] Miri Kazes,[‡] Kamalakannan Ranganathan,[†] Lothar Houben,[§] Dan Oron[‡] and Ariel Ismach[†*]

[†]Department of Materials Science and Engineering, Tel Aviv University, Ramat Aviv, Tel Aviv, 6997801, Israel
[‡]Department of Physics of Complex Systems and [§]Department of Chemical Research Support, Weizmann Institute of Science, Rehovot 7610001, Israel
[#]These authors contributed equally to this work
[*]aismach@tauex.tau.ac.il
Ariel Ismach,
Dept. of Materials Science and Engineering, Faculty of Engineering,
Tel-Aviv University Ramat-Aviv 69978, Tel-Aviv, Israel
Tel: +972-36405079

**Abstract**

Metal organic chemical vapor deposition (MOCVD) is one of the main methodologies used for thin film fabrication in the semiconductor industry today and is considered one of the most promising routes to achieve large-scale and high-quality 2D transition metal dichalcogenides (TMDCs). However, if not taken special measures, MOCVD suffers from some serious drawbacks, such as small domain size and carbon contamination, resulting in poor optical and crystal quality, which may inhibit its implementation for the large-scale fabrication of atomic-thin semiconductors. Here we present a Growth-Etch MOCVD (GE-MOCVD) methodology, in which a small amount of water vapor is introduced during the growth, while the precursors are delivered in pulses. The evolution of the growth as a function of the amount of water vapor, the number and type of cycles and the gas composition is described. We show a significant domain size increase is achieved relative to our conventional process. The improved crystal quality of WS$_2$ (and WSe$_2$) domains was demonstrated by means of Raman spectroscopy, photoluminescence (PL) spectroscopy and HRTEM studies. Moreover, time-resolved PL studies show very long exciton lifetimes, comparable to those observed in mechanically exfoliated flakes. Thus, the GE-MOCVD approach presented here may facilitate their integration into a wide range of applications.

**Keywords**

Metal-organic chemical vapor deposition, 2D materials, transition metal dichalcogenides, crystallinity, time-resolved photoluminescence, domain size

2D materials offer many interesting properties to be used in novel and existing technologies. One of the major bottle-necks in their successful integration in real-life applications is the need for reproducible large-scale growth of a high quality material. The pioneering work on the vapor-phase growth of single- and few-layer MoS$_2$[1] led to significant effort by the scientific community to realize methodologies for the large-area growth of 2D materials in general. As a result, extensive research has been carried to study chemical vapor deposition (CVD) growth of single- and few-layer atomic-films in general and transition metal dichalcogenides (TMDCs) in particular.[1-9] CVD of TMDCs often relies on the evaporation of metal oxide solid powders as the source for the transition metals (Mo, W, *etc.*) to be reacted with the chalcogen, usually obtained from a solid source as well. Despite the large volume of work and the great advance since then, metal-oxide based CVD growth has some significant drawbacks that inhibit its implementation for the study of the growth mechanism of such materials and for the consistent and rational growth of large-area 2D atomic-layers with the desired thickness and chemical composition. One of the main disadvantages is that the metal oxide precursors (MoO$_3$, WO$_3$, *etc.*) have relatively low vapor pressure, which forces their placement in proximity to the growth substrate *inside* the reactor chamber.[1, 6, 10] This is an unwanted situation when depositing thin films using CVD,[11] as there is little control over the delivery of the metal precursor to the growth substrate. Furthermore, the metal oxide reacts with the chalcogen as well, thus causing undesirable and uncontrolled changes in the metal vapor flux during the growth.[6, 10] In addition, there is a strong effect of the position of the precursor on the growth, and thus, several different materials are often simultaneously grown within relatively small areas.[6, 10] Indeed, interesting attempts were developed in order to overcome these drawbacks, such as the re-oxidation of the sulfurized metal oxide precursor during growth or the asymmetric CVD system configuration,[12] the growth on a catalytic substrate[9] and the growth in confined spaces.[13] However, these partial solutions cannot lead to a consistent and well-controlled growth of single- and few- layer 2D materials to be used in large-scale applications. Moreover, the possibility to add dopants or alloying elements in a controlled manner, which is highly desirable for a wide range of applications, is very limited as well.

Indeed, learning from the vast thin film research in the past few decades, in particular from the chemical vapor deposition of Mo- and W-based thin films,[14, 15] volatile metal precursors were studied as well.[3, 16-24] In this scheme, metal-carbonyls and -halides are among the most promising precursors. Park *et. al.* pioneered the metal-organic CVD (MOCVD) of TMDCs using molybdenum- and tungsten-hexacarbonyl for the synthesis of wafer-scale $MoS_2$ and $WS_2$ single-layer films. Taking advantage of the high-vapor pressure of such precursors, they are placed in bubblers *outside* the growth chamber, thus providing two main improvements: *i.* The precursor delivery to the growth area can be controlled by the carrier gas flow. *ii.* The growth of complex heterostructures can be achieved by changing the precursor flow *during* the growth.[25] This means that *in-situ* doping, alloying and heterostructure formation schemes could, in principle, be implemented as well in a controlled manner. However, the use of these volatile precursors was found to have some significant drawbacks. First, in these pioneering works, the growth rate was very slow, a continuous single layer TMDC required about ~26 hrs growth time. Second, the domain size achieved was significantly smaller (usually from few 10s to 100s of nanometers)[16, 24] than the one obtained using the metal – oxide precursors (~100 microns).[6, 8-10]

On the other hand, the electrical and optical quality of most of the CVD-derived 2D TMDCs semiconductors, have often shown to be poorer compared to that of mechanically exfoliated layers form 3D crystals,[26] possible due to defects creation in such high temperature processing and their exposure to air environment. Although several chemical treatment methods were developed to restore the intrinsic properties of 2D materials,[27-29] implementing such chemical passivation during large-scale processing of 2D semiconductors is challenging. Therefore, despite the extensive research and advance, a reproducible and large-scale growth methodology to produce atomic-layers with high crystal quality is still needed.

In order to improve the CVD (MOCVD)-derived TMDC crystallinity and thus its electrical and optical properties, we refer to three main approaches. The first method relies on the nucleation density reduction/control to obtain larger domains. This was achieved using halide salts (NaCl, KI, *etc.*)[30, 31] and the so called, "seed-promoters".[7] In these reports, the domain size was significantly increased to dimensions comparable with those obtained by the metal-oxide approach. However, these seed-suppressing compounds are placed *inside* the growth chamber as well, and thus, some of the inhomogeneities arising from the distance dependence of the diffusion to/on the growth substrate are relevant here as well. Furthermore, some of these compounds were found to have detrimental effects on the electrical properties of as grown TMDCs.[32, 33] Hence, these approaches may not be suitable for large-scale applications. The second path to improve crystallinity is by the seeded-growth approach, in which well-defined metal (or other) seeds catalyze the nucleation and growth.[8, 34] However, some of the seed material may affect in diverse ways the growth and the resulting optical properties of the grown material.[8] In addition, implementing this methodology in MOCVD might be difficult due to the very high nucleation density, inhibiting the selective nucleation at the seed. The third approach to improve the crystallinity of the film is to achieve epitaxial growth, usually referred as van der Waals epitaxy in layered materials.[35] The basic idea is that the domain size and nucleation density are not as important since the growing domains are well aligned, forming a highly crystalline (possibly single-crystal) film by the coalescence of isolated domains. This was reported in some TMDCs grown on sapphire,[3, 31, 36, 37] GaN[38] and also on other 2D materials such as graphene and different TMDCs.[39-42] Epitaxial growth is however, not entirely understood, despite promising reports attempting to do so.[36] Another problem in the growth of such atomic films is that, due to the complexity of the precursor flux delivery and the large-number of chemical reactions involved, it is difficult to model and understand the kinetic processes. Obviously, the different chemistries arising from the various precursor types (metal-organics, -chlorides and -oxides) have a significant impact on the nucleation and growth of the TMDC films. This variation in kinetics strongly impacts growth parameters, such as the domain size (100s of microns for the metal oxide-based CVD and ~100s of nanometers for MOCVD). Depending on precursor type, many complex processes may occur simultaneously, such that chemical reaction rates change during the growth due to local concentration variations, possible surface modification of the substrate, domains and domain-edges.[18, 43-47] All these will dictate eventually the growth type, lateral (2D) *versus* vertical (3D), and the maximum domain size achievable.

Here we present a "Growth-Etch-MOCVD" (GE-MOCVD) approach to improve the crystallinity of single-layer $WS_2$ (and $WSe_2$) layers. The original high-density nucleation is suppressed by the pulsed-delivery of the metal and chalcogen precursors in a constant flow of very low amounts of reactive species ($H_2O$) in between the growth-cycles. Thus, carbon contaminants as well as part of the TMDC nuclei are etched and/or re-evaporated, leading to the formation of larger domains with improved crystallinity. We can show a significant increment in the domain size, from few nanometers to more than 15 microns. The growth evolution of atomic–layers is studied as a function of the $H_2O$ vapor flow, gas composition and cycle number and type. The crystallinity of the GE-MOCVD grown TMDCs is compared to that produced by the standard MOCVD growth processes using Raman and photoluminescence (PL) spectroscopy, as well as high-resolution transmission electron microscope (HRTEM) studies. Furthermore, transient PL measurements show a long luminescence lifetime, comparable to the emission lifetimes measured from high-quality exfoliated flakes, suggesting high optical quality of the $WS_2$ films obtained by the GE-MOCVD approach. This methodology could be implemented in the MOCVD growth of many other 2D materials to improve their crystal quality, as required in nanoelectronics and optoelectronics.

## Results and Discussion

Figures 1 (a)-(c) show the results for *our* standard MOCVD growth of $WS_2$ films, using $W(CO)_6$ and Di-tert-butyl sulfide (DTBS) as the metal and chalcogen precursors, respectively. Here by "standard" MOCVD we refer to a process in which there is a continuous flow of the precursors and no water vapor is added. The schematic illustration of the customized MOCVD system is depicted in Figure S1. Figures 1(a)-(c) and S2 show typical results of such "standard" MOCVD growth type. Continuous layers of $WS_2$ were readily grown and transferred to arbitrary substrates, see experimental section and SI for details. Figure 1(a) shows an optical micrograph of a continuous layer grown on $SiO_2$(280nm)/Si substrate, which appears non-homogeneous, presumably with ad-layers (darker regions) as confirmed by AFM topography maps (Figure S2(a)). Raman spectroscopy of such samples exhibits the characteristic $WS_2$ modes 2LA(M), $E^1_{2g}(\Gamma)$ and $A_{1g}(\Gamma)$ modes, Figure S2(b). Very often, amorphous carbon (a-C) is detected in addition to the $WS_2$ peaks, as also seen in Figures 1(f) and S2(b), most probably, carbon contamination is from both, the organo-metal ($W(CO)_6$) and organo-sulfur (DTBS) precursors. We note here that the a-C contamination can be reduced to some extent by proper selection of precursors and process parameters, for example DTBS can be replaced by H$_2$S gas, which is a carbon-free source.[48] In our process however, carbon contamination could not be avoided completely, in contrast to some previous reports, probably due to differences in the system design.[3, 48, 49] Figures S2(b)-(c) show the Raman spectra of WS$_2$ atomic-layers on sapphire grown in standard MOCVD process using H$_2$S as a sulfur precursor. Although significant amount of a-C is reduced as compared to that of the DTBS precursor, considerable amount of a-C is still observable from the Raman spectra (Figure S2(c)), indicating that the a-C from metal organic (W(CO)$_6$) precursor is hard to avoid in our standard MOCVD process. In fact, theoretical and experimental validation in the past, also showed the a-C contamination when using organo-metal and H$_2$S as precursors.[43] This is one of the key points requiring special care when using such metal-organic precursors in CVD.[11] The residual amounts of a-C may not show significant effects on the growth of the conventional thin film materials, however, it is very critical for controlling the growth of three-atom-thick semiconductors like 2D TMDCs. Raman spectroscopy characterization of these a-C contaminated films caused damage, even at very low laser powers, as shown in Figures S2 (d) and (e), suggesting poor crystal quality.

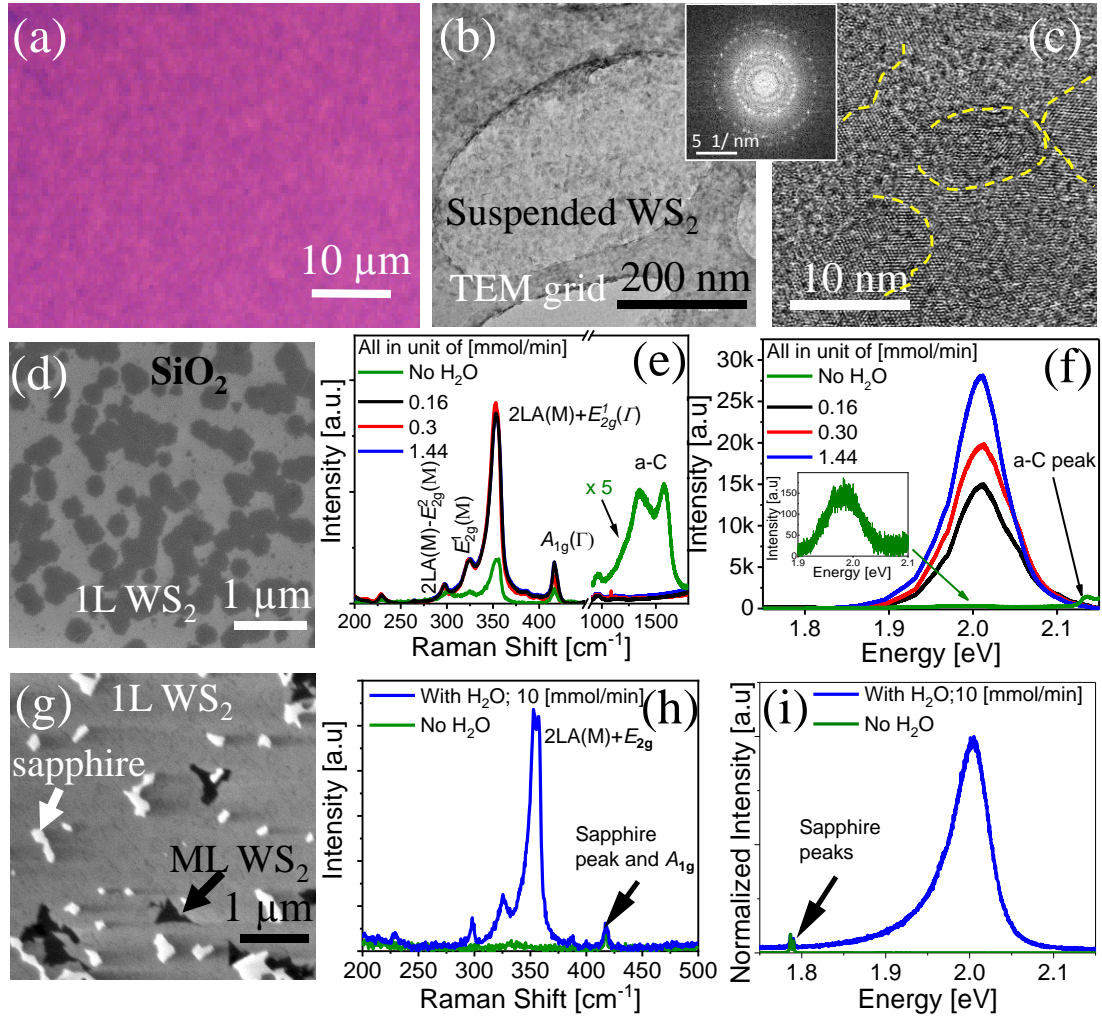

**Figure 1: MOCVD growth of WS$_2$ atomic-layers grown with and without H$_2$O vapor.** (a) – (c) Results obtained for a "standard growth" with no water vapor. Typical optical micrograph (a), low and high resolution TEM micrographs, (b) and (c), of continuous WS$_2$ atomic-layers grown on SiO$_2$/Si substrate. The yellow dotted lines in (c) denote the nanocrystalline domains of the continuous film. The inset shows the FFT image recorded from the lattice image shown in (c), see Figure S2. (d)-(f) Water-vapor assisted MOCVD growth of WS$_2$ layers on SiO$_2$/Si substrate. Typical SEM micrograph (d), comparison of Raman (e), and PL (f) spectra recorded from the samples grown under various concentrations of H$_2$O. The characteristic Raman modes of a-C are emphasized in (e) and (f). (g)-(i) The same for an H$_2$O-assisted MOCVD growth on *C*-plane sapphire.

Figure 1(b) shows a low-magnification TEM image of WS$_2$ transferred to a holey carbon TEM grid. Figure 1(c) displays an HRTEM image of the suspended layer region, in which the lattice of a nanocrystalline (few 10s of nanometers domain size, yellow dashed lines) film is observed. The FFT of the image in (c), inset, clearly shows a polycrystalline pattern, supporting the nanocrystalline nature of the WS$_2$ film. This agrees well with reports on MOCVD grown TMDC layers, where domain sizes of up to ~200 nm were observed.[16-18, 48] Larger domains in MOCVD processes were only obtained when adding metal halides to the system.[16, 31] Such unwanted carbon deposition have many detrimental effects, like serving as nucleation centers, thus increasing the nucleation density (and reducing the domain size), and general contamination of the growth substrate, surface and edges of the domains. In order to diminish this undesirable effect, low amounts of controlled water vapor (~300 ppm) were introduced into the growth chamber. This was achieved by adding N$_2$ gas from a relatively low purity source (99,99%) and a known humidity (and impurity in general) level. The amount of water in the N$_2$ gas was measured using a residual gas analyzer (RGA, Figure S3) prior to the growth, see the SI for further details. The addition of water vapor was reported in the past to affect the morphology of the grown domains as well as their crystal quality.[50, 51] In our case, it had an immediate and significant effect as shown in Figures 1(d)-(i). Figure 1(d) shows a SEM image of WS$_2$ growth on SiO$_2$/Si with the addition of 0.16 mmol/min of water. Analysis of the different Raman modes demonstrate the films are mostly monolayer in nature. The intensity ratio I(2LA(M)) /

I($A_{1g}$) is found to be ~3.3, confirming the 1L WS$_2$ nature (Figure S4).[52] The PL emission is centered around ~2.0 eV. The Raman and PL spectroscopy of WS$_2$ domains resulting from the growth under different water vapor flowrates are shown in Figures 1(e) and 1(f), respectively. In both cases, the peak intensity clearly grows with the increased H$_2$O flow.

Similar results were obtained on sapphire; Figures 1(g) and S4(b)-(c) show SEM images, highlighting the single-layer (1L), multi-layer (ML) and uncovered sapphire areas. The Raman (Figure 1(h)) and PL (Figure 1(i)) spectra of such samples exhibit higher intensities when the H$_2$O was added to the growth. The characteristic Raman signature of amorphous carbon detected in the water-free samples disappeared upon the addition of H$_2$O to the growth process, Figures 1(f) and S2(b)-(c). The lack of detectable a-C and the significant increase in the Raman and PL intensities are a clear indication for the improved crystallinity and purity of the layers. Such an effect can be explained by the reaction of H$_2$O with the a-C deposited on the sample and its re-evaporation, cleaning both the growth substrate uncovered areas and the domains. Similar results were obtained for WSe$_2$ atomic-layers (Figure S5(a)-(f)), thus suggesting its generality for the MOCVD growth of TMDs. However, such growth methodology did not change significantly the size of the domains (of WS$_2$ and WSe$_2$), as it is limited by the presence of surface impurities as well as the partial pressures of the metal and chalcogen precursors, which does not change during the growth. This suggests there is a limit to the film improvement with the water vapor addition approach. Carbon contamination may enhance the non-radiative recombination process and thus the PL intensity increase with the water flowrate. As mentioned above, based on the results summarized in Figures 1(a)-(c) and S2, as well as previous reports,[3, 16-18, 32, 48] there is a high nucleation density when using metal-organic precursors. The W(CO)$_6$ is known to thermally decompose at relatively low temperatures,[53] compared to the ones used here and needed for the growth of highly crystalline TMDCs. This implies a very fast metal-precursor decomposition upon its introduction to the pre-heated reactor. This, together with the carbon contamination, from both precursors, could be the cause for the high nucleation density usually obtained in MOCVD growth of TMDCs. Therefore, there is a need to partially suppress the nucleation centers in order to allow the domains to expand and grow over larger areas. To achieve this goal, and based on the encouraging growth results with a low flowrate of water vapor, a pulsed-growth approach was implemented. In this scheme, the metal and chalcogen precursors were supplied to the system in *pulses* (the "growth" step), while the Ar and water vapor were flowed continuously (the "etch" step).

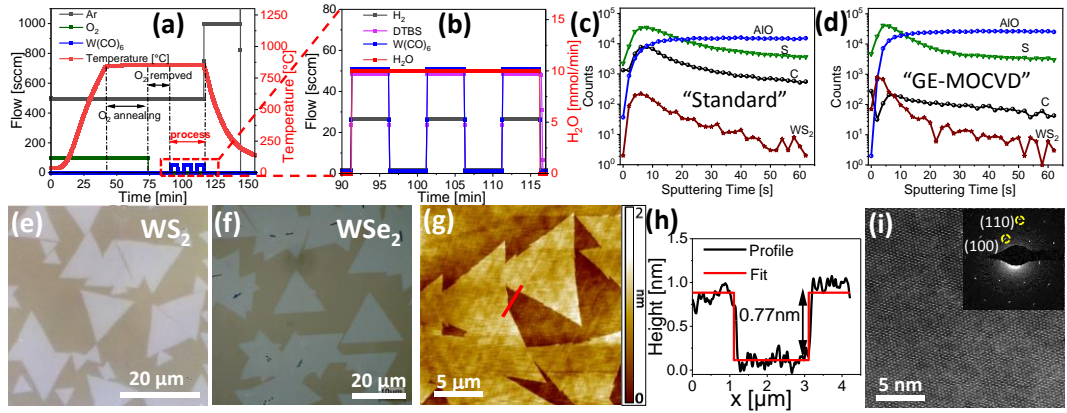

**Figure 2: Growth-Etch MOCVD Approach.** (a), (b) Flow *vs*. time charts for the GE-MOCVD growth. TOF-SIMS depth profiles, counts as a function of sputtering time, recorded from "standard" MOCVD (c), and GE-MOCVD (d) grown samples. The decrease in the amount of carbon in (d) is clearly seen. The obtained high-quality atomic-layers; Typical optical micrographs of WS$_2$ (e), and WSe$_2$ (f) grown on *c*-sapphire. Topography map (g), and the corresponding height profile drawn along the red line is shown in (h). (i) Atomic-resolution HAADF STEM micrograph of a WS$_2$ single crystalline domain. Inset shows the SAED pattern recorded from a single crystalline WS$_2$ domain, see SI for details.

Figure 2 shows the results obtained by the GE-MOCVD approach, in which the gas flow chart as a function of time is depicted in (a)-(b). The precursor pulses and the continuous supply of H$_2$O are shown in Figure 2(b). The Raman spectroscopy characterization clearly showed the a-C peaks disappear upon water addition to the MOCVD growth (Figures 1(d)-(i)). In order to further verify the reduction in the amount of carbon on the growth substrate upon implementation of the GE-MOCVD approach, time-of-flight secondary ion mass spectroscopy (TOF-SIMS) depth profiles were recorded on "standard" MOCVD and GE-MOCVD grown samples. Here, we considered WS$_2$ samples grown in "standard" MOCVD process using carbon-free H$_2$S as a sulfur precursor to compare with the samples grown in GE-MOCVD process using high carbon source of DTBS as the sulfur precursor. Figures 2(c)-(d) show that almost two orders of magnitude higher amounts of a-C are detected on the "standard" MOCVD sample (Figure 2(c) and Figures S6(a),(b)), as compared to that of the GE-MOCVD sample (Figure 2(d) and Figures S6(c),(d)). It is also evident from the TOF-SIMS measurements that, even an metal-organic (W(CO)$_6$) precursor alone can incorporate significant amount of carbon contamination in $WS_2$ monolayers grown in the "standard" (*our standard*) MOCVD process. Whereas, GE-MOCVD approach showed that, it can effectively reduce the a-C contamination even when high-carbon sources of chalcogen precursors like DTBS are used, along with the $W(CO)_6$ precursor. The typical optical micrographs of $WS_2$ and $WSe_2$ (Figures 2(e)-(f)), show a significant increase in the domain size. The AFM topography map of the pristine $WS_2$ atomic layers shows the monolayer domains of thickness ~ 0.77 nm (Figures 2(g)-(h)). The atomic-resolution HAADF STEM micrographs along with SAED data, inset, show the high crystallinity of the domains (Figures 2(i), S7(a)-(c)). The microscopy characterization demonstrates the significant improvement in the $WS_2$ crystal quality grown by the GE-MOCVD approach. As discussed above, the water vapor reacts with the solid carbon contaminants to produce CO gas (and maybe $CO_2$), causing their re-evaporation and therefore cleaning the substrate and domains. The main sources for the carbon contamination are the $W(CO)_6$ and the DTBS,[11, 48] therefore, addition of water vapor to a "standard" MOCVD growth alone has a limited capability to eliminate such contaminants. This is due to the fact that the precursors are supplied continuously, and hence, the a-C is deposited throughout the growth process. In contrast, a pulsed-growth method of GE-MOCVD approach, in which the precursors ($W(CO)_6$ and DTBS) are supplied in cycles, but the water vapor (and carrier gases) is supplied continuously, allows for a successful re-evaporation of the amorphous carbon, thus cleaning the substrate and domains during the precursor-free pulse, before the growth is continued. This approach significantly improved the domain size and crystal quality. The absence of the a-C Raman modes, the low amount of carbon species on the GE-MOCVD grown samples and the increase of the crystalline domain sizes, confirm the GE-approach effectively removes the majority of carbon contamination arising from all the precursors.

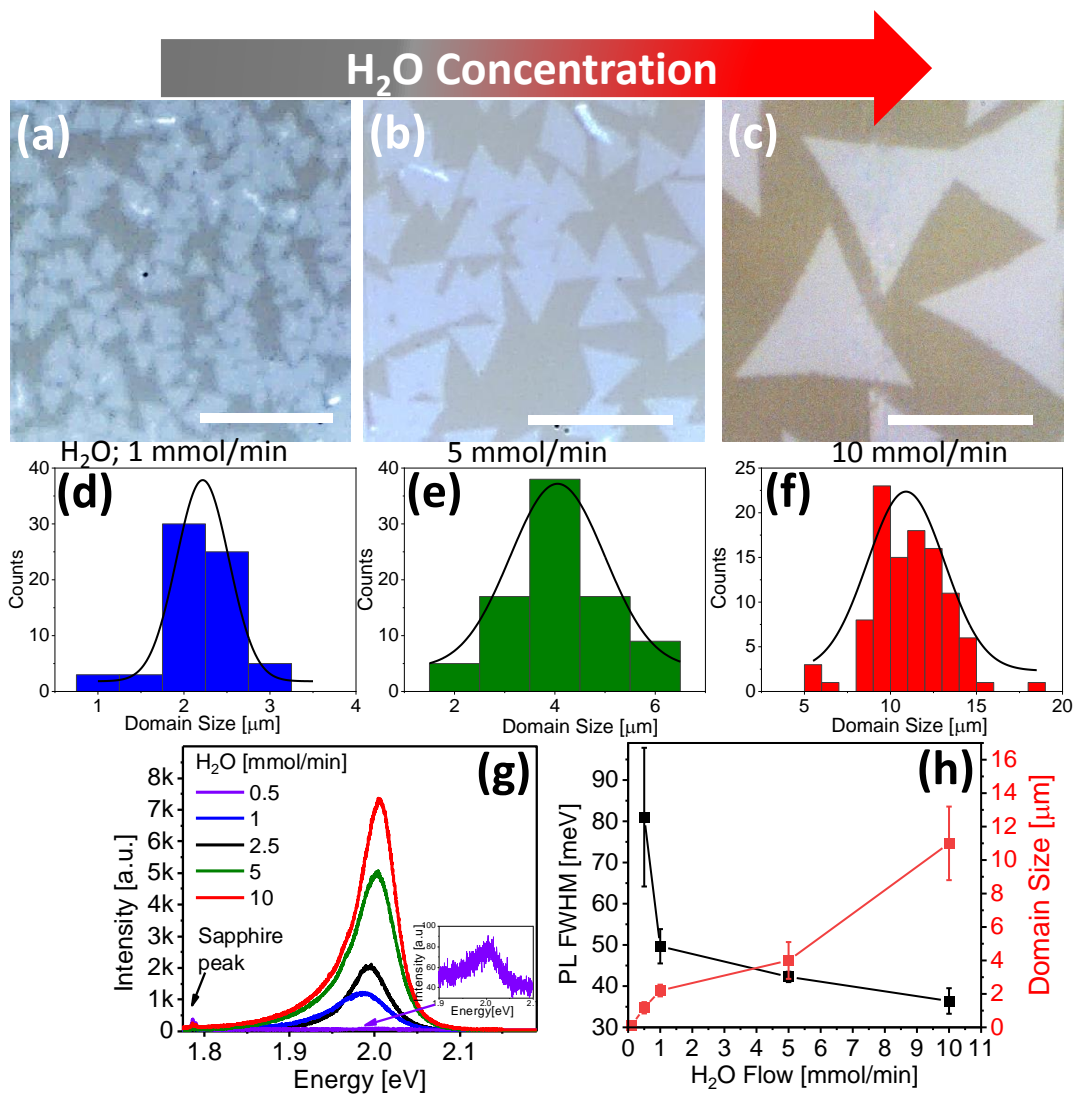

**Figure 3: The effect of $H_2O$ concentration in the GE-MOCVD growth**. (a)-(c) Typical optical micrographs of $WS_2$ atomic-layers grown under increased concentration of $H_2O$; 1, 5, 10 mmol/min. The scale bar is 10 µm. (d)-(f) The corresponding domain size distributions. (g) The comparison of the PL spectra recorded from $WS_2$ domains grown under different concentrations of $H_2O$. (h) PL FWHM and domain size as a function of the $H_2O$ concentration, showing a clear domain size evolution with the increased $H_2O$ vapor flow rates.

The effect of the $H_2O$ flowrate on the growth is summarized in Figure 3. Figures 3 (a)-(c) show optical images depicting the domain size evolution as a function of $H_2O$ flow, from ~0.1 to 10 mmol/min. The respective domain size distribution is shown in Figures 3 (d)-(f). The PL emission from such samples is plotted in Figure 3(g), in which it is evident that the PL emission intensity increases and full width half-maximum (FWHM) decreases with increased $H_2O$ flow, Figure 3(h), while the domain size increases by more than two orders of magnitude, from ~0.09 ± 0.04µm to 11.5 ± 2.5 µm, Figure 3(h). The gradual PL emission enhancement, together with the visible increase in the domain size (h), are a clear indication that the optical quality and crystallinity are dramatically improved.

The variation in the coverage area and domain size with the type and number of cycles in GE-MOCVD is analyzed and presented in Figure 4. Here the #G and #E refers to the number of cycles in which the precursors ($W(CO)_6$ and DTBS, #G, "growth step") and the $N_2$ (#E - with $H_2O$, "etch step") were supplied, see the SI for details. The micrographs of $WS_2$ domains grown with different cycle types 1G, 1G + 1E, 2G + 1E, 3G + 2E, respectively, show the gradual increase in the domain size (Figures 4 (a)-(e)). Figures 4(e)-(f) display the characterization of the fully covered $WS_2$ monolayer obtained with a 4G + 3E cycle type, showing SEM micrograph (e), AFM topography (f), and an optical micrograph of a 2" sapphire wafer fully covered with an $WS_2$ monolayer, (g). The coverage and the domain size of the $WS_2$ atomic-layers as a function of the cycle number and type are plotted in Figure 4(h). Both parameters can be controlled *via* the cycle design in order to achieve large domain size of ~12 μm and coverage (100%).

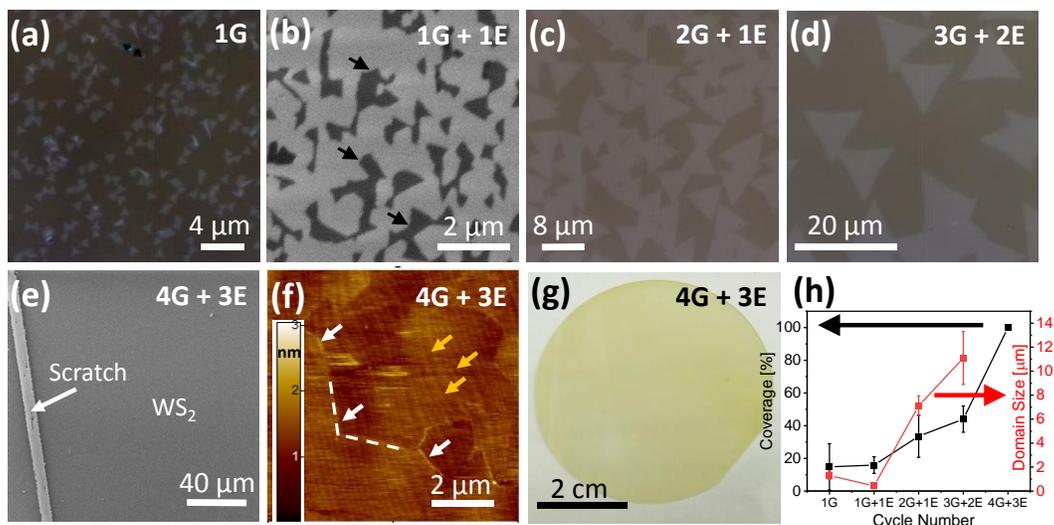

**Figure 4: The effect of the cycle # and type on the growth.** (a)-(f) Domain evolution as a function of cycle type. (a) Optical micrograph of the 1G growth cycle. (b) SEM image of the "1G + 1E" cycle, the $WS_2$ domains appear dark, black arrows. (c)-(d) Optical micrographs of the "2G + 1E" and "3G + 2E", respectively. (e) - (f) Full monolayer $WS_2$ obtained in a "4G + 3E" cycle, as observed in SEM, (e), AFM, (f), and optical microscope on a 2" sapphire wafer, (g). The yellow arrows in (f) mark the atomic steps of the C-plane sapphire and the white arrows, and dashed line, emphasize the $WS_2$ domain boundaries. Area coverage and domain size as a function of cycle # and type.

Figure S8 (a) shows the PL spectra obtained from a full layer of $WS_2$ sample grown with a 4G + 3E cycle type before (black) and after being transferred from sapphire to a $SiO_2$(280nm)/Si substrate (red), the inset shows the transferred film. The blue shift in PL peak of transferred layers is due the relaxation of strain generated during the growth.[27] In order to study the effect of $H_2$ flowrate on the cycle type for the growth and domain alignment with respect to the C-plane sapphire, we have modified the cycle 3G+2E in its last cycle. Figure S8 (b) shows the optical micrograph of the $WS_2$ domains obtained in a modified 3G + 2E cycle, in which the $H_2$ flow was doubled (from 25 sccm), in the last pulse. It can be shown that not only the domain size was significantly increased from sub-micron (1G cycle) to more than 10 microns (3G + 2E cycle), but also their relative orientation with the substrate (*C*-plane sapphire with a primary cut along *a*-axis) is more uniform, plotted in Figure S8 (c). Such result might be an indication of an improved substrate-layer interaction (for example by an $H_2$-based enhanced surface cleaning of contaminants). Another option could be the selective etching of misoriented domains, which are presumably less stable than the epi-grown layers. Further experimental research and theoretical calculations are needed to clarify this important task.

In many vapor-phase growth methods, the use of $H_2$ may help to react with residual oxygen and moisture present in such systems, avoiding by that unwanted reaction/etching processes with the grown material. In order to understand the hydrogen role on the TMDC formation, the growth evolution under different flowrates of $H_2$ is studied. The hydrogen effect on the growth with 1 cycle (1G) is summarized in Figure S9. The nucleation density and domain size as a function of the $H_2$ flow is plotted in Figure S9(e). A higher $H_2$ flow, 75 sccm, leads to very high nucleation densities and thus, smaller domain size. The nucleation density is reduced by around three orders of magnitude, with the $H_2$ flow below 50 sccm. On the other hand, an $H_2$-free three-cycle growth, Figure S10, leads to very low nucleation density and small dendritic domains. This could be due to an excess of $H_2O$-based etching of the $WS_2$ domains, suggesting that some $H_2$ is necessary to react with the $O_2$ species present in the system and prevent and slow down the almost complete $WS_2$ re-evaporation.

Here, we aim to explain the growth mechanism and the results described in the previous sections. The introduction of small amounts of $H_2O$ during the growth and the pulsed delivery of the growth precursors have two main effects; First, it eliminates or diminishes dramatically, the amount of carbon impurity in the films and on the growth substrate, leading to purer and higher quality $WS_2$ films. Since the presence of amorphous carbon on the substrate may act as a nucleation center, its elimination alone leads to smaller nucleation densities and thus, larger domains. Second, $H_2O$ was shown to etch and re-evaporate TMDCs,[50, 51, 54] therefore, small amounts of water vapor present in the system, react with some of the small/defective $WS_2$ nuclei/domains causing their re-evaporation, reducing by that the nucleation density as well. Thus, the most relevant reactions taking place during the "etch-step" in which the substrate, and the grown domains, are exposed to water vapor, but not to metal and chalcogen precursors, are the following:[54, 55]

(1) $C(s) + H_2O(g) \leftrightarrow CO(g) + H_2(g)$
(2) $WS_2(s) + 3H_2O(g) \leftrightarrow WO_3(s) + 2SO_2(g) + 6H_2(g)$
(3) $WS_2(s) + 6H_2O(g) \leftrightarrow WO_2(s) + 2SO_2(g) + 6H_2(g)$

These reactions lead to the re-evaporation of carbon contaminants, (1), and small or defective $WS_2$ domains, (2)-(3). The free energy of formation of the $WO_3$ and $WO_2$ phases during the growth, is depicted in Figure S11. Defective and small $WS_2$ domains are more reactive due to their high density of defects/edges. Hence, the introduction of low amounts of water, will preferentially react with such domains, through reactions (2) and (3). Once the metal-oxide phase is formed, it will readily sublimate due to its atomically-thin nature and the high growth temperature used. In fact, the vast majority of metal-oxide based CVD growth is based on the sublimation of bulk powders of the same phases.[1, 6-10, 12, 40, 42, 56] Furthermore, such W-O phases are known to further react with water to form more volatile phases, such as $WO_2(OH)_{2(g)}$.[51, 57, 58] Reactions (2)-(3) may cause the *local* concentration of $H_2$ on the surface to increase. Hence, increasing the $H_2$ flow, Figure S9, causes an increase in the $H_2$ concentration in the growth system and may cause the reactions above to be slowed down and eventually reversed, depending on the surface concentration of $H_2$. A schematic model summarizing the growth processes in a standard- and GE-MOCVD is shown in Figure 5. The top panel represents the "standard"-MOCVD growth in which high nucleation density (thus small domain size) and a-C contaminants are often obtained. The bottom panel in Figure 5 depicts the growth mechanism in the GE-MOCVD approach. The "growth" cycle is followed by an "etch" pulse, where the metal and chalcogen precursors flow is stopped. This causes the $H_2O$ species to react with the a-C contaminants and small/defective $WS_2$ ($WSe_2$) domains, causing their re-evaporation. Therefore, leading to the growth of high-quality TMDC layers.

The GE-MOCVD approach can be effectively implemented to reduce the a-C contamination even from the $W(CO)_6$ and to obtain high crystalline and larger single crystalline domains using other chalcogen precursors such as carbon-free $H_2S$. Figure S12 shows the demonstration of such results, in which $WS_2$ domains grown using $H_2S$ as a sulfur precursor in the GE-MOCVD approach with a cycle of 4G + 3E. When compared the domain sizes obtained in GE-MOCVD approach with that of available reports on MOCVD grown 2D TMDs, the present results show very promising even without use of metal salts and seed promoters like NaCl, KI, and PTAS *etc*. (Table S1).

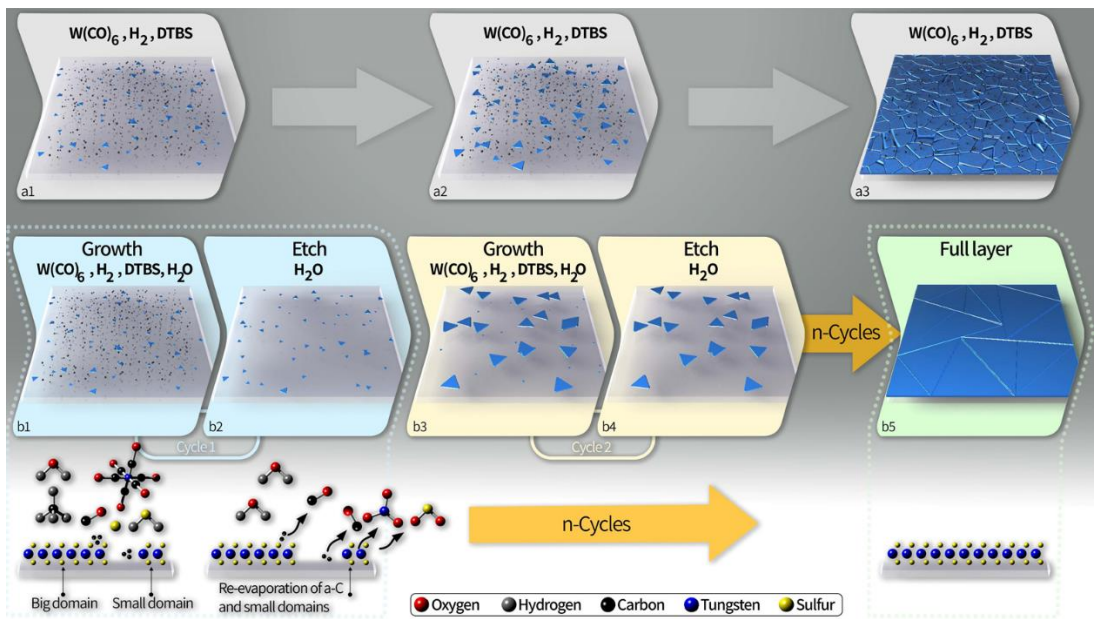

**Figure 5:** Schematic model of the growth mechanism in a "standard" MOCVD (top) and GE-MOCVD methodologies (bottom).

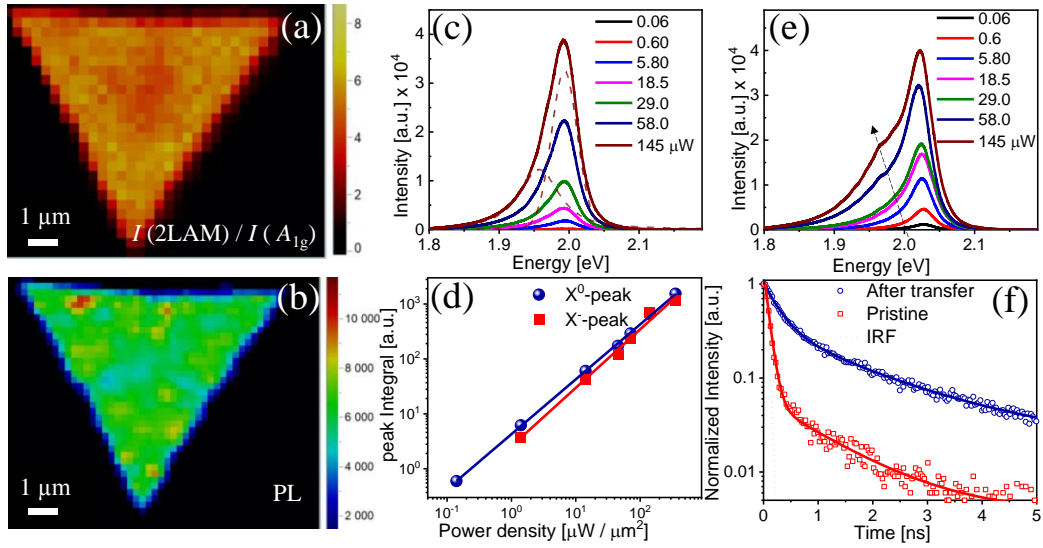

**Figure 6:** The spectroscopic maps of Raman intensity ratio of 2LAM and $A_{1g}$ modes (a), and PL intensity (b) of single crystalline WS$_2$ domain, showing uniform monolayer characteristics. (c) Steady-state power dependent PL spectra of pristine monolayer WS$_2$ domains. The dashed lines in (c) are the fitted curves corresponding to the highest excitation power spectrum. (d) The PL peak intensities of neutral-$A$ exciton ($X^0$) and trion ($X^-$) emissions from pristine domains as a function of excitation power. The solid lines are power law fits of PL intensity. (e) Steady-state power dependent PL spectra of transferred monolayer WS$_2$ domains. The dotted arrow shows the redshift of the low energy peak as a function of excitation power. (f) TRPL decay trace measured for a pristine and transferred single crystalline WS$_2$ domains. The solid lines are exponential decay fits of TRPL curves, and the IRF is plotted in a black dotted line.

Figure S13 shows single crystalline pristine (Figure S13 (a)) and transferred (Figure S13 (b)) WS$_2$ domains grown by the GE-MOCVD technique, in addition to the large area continuous atomic layers (Figure S13 (c)). The uniform monolayer characteristics of single crystalline WS$_2$ domain are verified by Raman and PL spectroscopic imaging. Figure S14 (a) shows the optical micrograph of isolated WS$_2$ domains that were used for such spectroscopic imaging. The Raman map (Figure 6(a)) generated by the intensity ratio of the 2*LAM* (Figure S14 (b)) and $A_{1g}$ (Figure S14(c)) phonon modes shows that the intensity ratio ($I$ (2*LAM*) / $I(A_{1g})$) is greater than 4, all over the domain, which is a characteristic nature of the single layer WS$_2$.[52] In addition, the homogeneous optical quality of the single crystalline WS$_2$ domain is evident from the uniform PL emission associated with the *A*-excitons, as seen in PL intensity (Figure 6(b)) and PL position (Figure S14 (d)) maps.

The optical quality of pristine and transferred WS$_2$ domains are further studied by conducting steady-state power dependent PL and time resolved PL (TRPL) measurements on single crystalline WS$_2$ domains. The PL spectra as a function of excitation power recorded from pristine WS$_2$ single crystalline domains show an emission around 2.0 eV (Figure 6(c)). The intensity of the PL emission increases as a function of increased excitation power without any appreciable shift in the peak position, indicating no heating effect on the sample due to excitation powers that were used for the measurement. A slight asymmetric broadening on the low energy side of the peak is also seen as the excitation power increases. The strong luminescence around 2.0 eV, can be deconvolved into two peaks. The peak centred at ~1.99 eV is identified as the neutral *A*-exciton ($X^0$) emission and the low energy peak around 1.96 eV is attributed to electron-bound exciton (negative trion; $X^-$) emission.[8] The neutral exciton emission is stronger than the trion emission at all incident excitation powers. The power (*P*) dependent PL integral intensities (*I*) of both emission peaks are plotted in log-log plot (Figure 6(d)) and fitted with power law, $I = P^b$, where, b is the slope. Both the $X^0$ (b = 0.99) and $X^-$ (b = 1.06) emissions show linear behaviour, confirming that there is no trap-assisted recombination through mid-gap states.[28, 59-61]

The transferred WS$_2$ single crystalline domain PL spectra shows an emission around 2.02 eV at low excitation powers (Figure 6(e)). The intensity of the PL emission increases as a function of increased excitation power without any appreciable shift in the peak center. However, the broadening and intensity of the low energy tail is also enhanced with increased excitation power and evolved into a clear peak centered at ~1.95 eV. Deconvoluted PL spectra presented in Figures S15 (a) - (b) show that the neutral *A*-exciton ($X^0$) emission peak is centered at 2.02 eV and is blue shifted (30 meV) as compared to that of the pristine sample, possibly due to strain relaxation following the transfer from the growth substrate.[27] Notably, no appreciable shift of $X^0$ emission is observed with increased excitation power. However, the low energy trion emissuchsion ($X^-$) peak has red shifted (Figures 6(e), S15(a) - (b) ) from 2.0 to 1.97 eV with increased excitation power, indicating the emission is caused by electron-bound exciton (trion) recombination.[27, 62, 63] The trion nature of the low energy peak ~1.97 eV is further confirmed by excitation dependent ratio of PL intensity of neutral exciton ($X^0$) to electron-bound exciton ($X^-$) emissions (Figure S15(c)). The ratio is increased with excitation power for both pristine and transferred samples. At low powers, both samples showed almost similar values of exciton to trion ratio. However, at high powers the transferred sample showed twice the ratio ($X^-/X^0$) as compared to that of pristine sample, indicating more free charge carriers bounded with excitons. The relative PL quantum efficiency (PLQE), which is the ratio of emitted PL to incident excitation power, is also an indicative of the optical quality of the samples under comparison. At the lowest excitation intensity, the PLQE is almost two orders higher for transferred $WS_2$ domains as compared to that of pristine domains (Figure S15(d)), plausibly due to the relaxation of substrate induced strain effects on the $WS_2$ domains during the growth.[27] As the excitation power increases, the PLQE decreased for transferred samples and approached values similar to those of pristine domains at the highest excitation power. The reduction in PLQE at higher powers might be due to the larger trion formation rate by the generation of free charge carriers and the intrinsically low PLQE of trions.[29]

The temperature dependent PL measurements can further shine light on the optical quality of the $WS_2$ monolayer domains. A distinct PL emission due to the intrinsic point defects that are formed during the growth in TMDs monolayers, can be clearly resolved at sufficiently low temperature.[60, 64-66] In general, the predominant intrinsic point defects in TMDs monolayers are found to be the chalcogen vacancies which are responsible for mid-gap states resulting in a defect emission around 250-300 meV below the emission due to neutral and charged excitons.[60, 64-66] In addition, the extrinsic defects formed by the adsorbed molecules such as $H_2O$, $O_2$ *etc*. on TMDs monolayers can degrade the optical quality of the samples further.[60, 64-66] In order to identify any intrinsic defects in monolayer $WS_2$ domains, temperature dependent PL measurements were performed on GE-MOCVD grown $WS_2$ domains on sapphire. Figure S15(e) shows that as the temperature decreased from 300 K to 80 K, no significant defect emission is absorbed below the emission due to neutral and charged excitons, confirming the high optical quality of the domains. The excitation power dependent PL at 80 K (Figure S15(f)) also shows no significant change in the peak shape of the exciton emission even at high pump power, revealing the insignificant effect of the extrinsic defects on WS$_2$ monolayers.[65-67] The low defect density in GE-MOCVD grown WS$_2$ monolayers is further evident from the Raman spectroscopic measurements. The longitudinal acoustic (LA) phonons with momentum q ≠ 0 at *M* point of the Brillouin zone (LA(M)) are found to originate from structural defects mostly sulphur vacancies (V$_S$).[68, 69] The intensity ratio of LA(M) and first-order $E^1_{2g}(\Gamma)$ modes, I(LA) / I($E^1_{2g}$) is an indication of the crystal quaity and it can be used to compare the values among the different samples grown in MOCVD. From Figure S16(a), it can be clearly observed that the intensity ratio is much smaller in case of GE-MOCVD ( for DTBS = 0.19, for H$_2$S=0.22) approach as compared to the conventional MOCVD method (0.44), suggesting the presence of less number of defect states, and hence the improvement in the crystallinity.

The high optical quality of the single crystalline WS$_2$ domains grown by GE-MOCVD technique is further confirmed by studying the exciton dynamics using time-resolved PL (TRPL). Figure 6(f) shows the TRPL spectra recorded from single crystalline WS$_2$ domains of pristine and transferred samples, at a pump fluence of ~5 nJ·cm$^{-2}$. The normalized TRPL signal of pristine domains is fitted with bi-exponential decay having a fast decay time constant ($\tau_1$) of ~ 85 ps and a slow decay time constant ($\tau_2$) of ~ 1.51 ns with average lifetimes ($\tau_{avg}$) of 0.733 ns. TRPL of the transferred WS$_2$ domains also fitted with bi-exponential decay having a fast decay time ( $\tau_1$) of ~ 270 ps and a slow decay time of $\tau_2$ ~ 1.47 ns with an average lifetimes ($\tau_{avg}$) of 1.15 ns. We note that for the transferred WS$_2$ an additional long lived lifetime component (few nsec) was observed but its magnitude was small and for the sake of clarity will not be discussed here further. Here, we fitted TRPL data with bi-exponential decay to study any nonradiative transitions associated with intrinsic defects along with radiative recombination of excitons. The fast decay component ($\tau_1$) is attributed to the band edge emission typically shown in TMDCs to be suppressed by the onset of nonradioactive recombination through mid-gap states formed by defects, so called trap assisted recombination mechanism.[70-72] The slow decay time ($\tau_2$) is assigned to radiative recombination of the excitons.[73] Longer decay times of $\tau_1$ means lesser the nonradiative recombinations associated with intrinsic defects. Note that, the nonradiative recombination of excitons through exciton-exciton annihilation (EEA) process also reduces the exciton life times, however, such EEA process happens only at sufficiently higher pump fluency, typically >100 nJ.cm$^{-2}$.[70-72] Therefore, in order to study the presence of intrinsic defects and compare the optical quality of the GE-MOCVD grown monolayers of WS$_2$ with that of the high quality mechanically exfoliated samples, we measured the TRPL spectra at very low pump fluence in which the density-independent decay channels (both defect-mediated and defect-free radiative recombination) dominates and EEA process is insignificant (Figure S15 (b)). The $\tau_1$ for the transferred WS$_2$ domains shows 3 times longer as compared to that of pristine domains, indicating the relaxation of strain induced effects by growth substrate and formation of fewer mid-gap states, therefore consistent with higher PLQE as measured in steady-state PL measurements (Figure S15(d)) and with the presence of a longer-lived emission component. When compared both nonradiative and radiative lifetimes of transferred GE-MOCVD grown WS$_2$ domains with that of the reported exfoliated samples (Table S2), GE-MOCVD samples showed slightly higher values ($\tau_1$, $\tau_2$, and $\tau_{avg}$ ≈ 0.3, 1.6, and 1.22 ns at 0.8 nJ.cm$^{-2}$) than that of exfoliated samples ($\tau_1$, $\tau_2$, and $\tau_{avg}$ ≈ 0.23, 1.33, and 0.55 ns at 1.3 nJ.cm$^{-2}$ for bi-exponential fit; and $\tau_1$ ≈ 0.806 ns at 5 nJ.cm$^{-2}$ for a single exponential fit)[71, 72] Such a considerable slowing down of nonradiative recombination, highlighting the presence of significantly fewer defect states, further confirm the high optical quality of the WS$_2$ atomic layers grown by the GE-MOCVD technique.

**Conclusions**

In summary, a Growth-Etch MOCVD is developed for the synthesis of high optical quality WS$_2$ (and WSe$_2$) atomic layers. The carbon contamination often present when using metal-organic and the organo-sulfide (and selenide) precursors in standard MOCVD is eliminated *via* a sequential growth and etch process in which the precursors are delivered to the system in pulses and the continuous flow of low amounts of water vapor. As a result, the ultra-high nucleation density usually obtained in standard MOCVD is highly suppressed by the re-evaporation of the carbon contaminants, and small and defective TMDC nucleus. Therefore, this methodology leads to improved crystallinity and lower contamination, as confirmed by TOF-SIMS characterization. By understanding the growth evolution in the developed methodology, high crystalline quality TMDCs were achieved in relative short times, as required for industrial applications. The Raman, steady-state and time-resolved PL spectroscopy studies established the improved crystal and optical quality of the atomic layers grown by the GE-MOCVD. The results obtained here clearly show that these films have very promising optical quality for a CVD (or MOCVD)-derived TMDC, comparable to the ones obtained from mechanical exfoliation of a high quality 3D crystal. Hence, the GE-MOCVD method has the potential to address the more acute deficiencies in MOCVD growth of TMDCs, such as the a-C contamination, low growth rate and poor crystallinity. This procedure could be implemented for the synthesis of many other 2D materials and as a proof of concept, the results on $WSe_2$ showed here. The large-scale growth of high-quality atomic-films is a prerequisite for their integration into a wide range of applications. The work presented here therefore, is a significant step forward for this to happen.

## Supporting Information

This material is available free of charge at The Supporting Information is available free of charge on the ACS Publications website at DOI:

Schematic of MOCVD setup; Calculation of the precursors and water flowrates; additional Raman, PL, TRPL, TEM, and TOF-SIMS results; and tables corresponding to the comparison of present results with literature.

## Acknowledgements

The authors gratefully acknowledge the very generous support from the Israel Science Foundation, projects # 2171/17 (A.C. and P.K.M.), 2549/17 (A.P.) and 1784/15 (A.I.). Authors acknowledge the expertise of Dr. Alexander Gladkikh for TOF-SIMS analysis.

## Experimental

### MOCVD and GE-MOCVD Growth

The synthesis of $WS_2$ was carried out using a hot wall 3 inch customized MOCVD furnace (CVD equipment corporation, model Easy Tube 2000), equipped with 4 separate bubbler for precursors. In one bubbler, Di-tert-butyl sulfide (DTBS, sigma Aldrich, 97%) was loaded inside a glove box under inert gas and the second bubbler was loaded with $W(CO)_6$ (Strem chemical, 99.9%) in the same glove box. The choice of other source of sulfur was $H_2S$. For this, we used a scrubber system to deal with the residual gas. The background and carrier gas used were argon (99.9999%) and hydrogen (99.999%). The substrates, $SiO_2$ and *c*-plane sapphire (annealed at 1050 °C for 10hrs) were cleaned using acetone and IPA each for 10 minutes in an ultrasonicator and were dried using a nitrogen gun. Prior to the growth, the furnace was evacuated to a pressure down to 15mTorr for 15 minutes to remove any unwanted moisture/oxygen species. Thereafter, the furnace was ramped to 850 °C at a heating rate of 20 °C/min for 30 minutes under 50 torr with 100sccm of oxygen to remove any possibilities of carbon contamination. Then the oxygen flow was stopped and the system was maintained at the same (850 °C) temperature for another 25 minutes. All the required precursors were then released for the growth. We followed a growth and etch technique in our process cycle as schematically shown in Figures 2 (a)-(b). The process cycle was mainly a combination of two main steps with a period of 5 minutes. In the first step, all the precursors ($H_2$, DTBS, $W(CO)_6$, $H_2O$) were introduced for the growth to take place. In the 2nd step, only $H_2O$ was introduced for the etching process. We repeated this sequence for 3-4 cycle, always terminating with the growth step. The hydrogen and $H_2O$ flow was varied from recipe to recipe. The amount of $H_2O$ introduced was measured by the (residual gas analyzer) RGA technique as described in the SI (Fig. S3). The calculated flow of $W(CO)_6$ and DTBS was found to be ~ $3.28 \cdot 10^{-7}$ mol/min and ~ $3.25 \cdot 10^{-4}$ mol/min respectively, which leads to a sulfur to metal ratio of (S/M) ~1000. The detailed calculations are in the SI.

The GE-MOCVD growths were carried out with the following parameters: Temperature 850 °C, pressure 50 torr, DTBS ~ $3.25 \cdot 10^{-4}$ mol/min, $W(CO)_6$ ~ $3.28 \cdot 10^{-7}$ mol/min, $H_2$ 25 sccm, Argon 500 sccm. The molar flow of $H_2O$, when relevant, is mentioned in each figure.

**Transfer of $WS_2$ to TEM grids, quartz, and $SiO_2$ substrates**

The transfer of $WS_2$ film from the growth substrate to a new desired substrate was carried out by using a previously reported polystyrene (PS) technique.[56] In this process, 450 mg of PS (280000 g/mol) was dissolved in 5 ml of toluene. The PS solution was then spin-coated

(60 seconds at 3500 rpm) onto the as-grown WS$_2$ layer on sapphire to make a thin film of PS. Thereafter, the sample was backed at 90 °C for 30 minutes and then at 120 °C for 10 minutes. The polymer/WS$_2$ assembly was delaminated by allowing the water to penetrate in between sapphire and the assembly. The floating layer was then fished out of the water to any desired substrate (TEM grid, quartz or SiO$_2$) and left for drying naturally at room temperature. Another baking process (90 °C for 30 minutes and more 15 minutes at 120 °C) was adopted to ensure no water residuals were left in the interface. In the end, the PS film was dissolved from the assembly by using toluene.

**HRTEM and HAADF measurements**

High-resolution scanning transmission electron microscopy (HRSTEM) images were recorded in a double aberration-corrected Themis Z microscope (Thermo Fisher Scientific Electron Microscopy Solutions, Hilsboro, USA) equipped with a high-brightness FEG at an accelerating voltage of 80kV. High-resolution TEM images were collected on the Themis-Z instrument on a Gatan One View CMOS camera (Gatan Inc., Pleasanton, USA) at negative spherical aberration settings (Cs=-20µm) with higher order axial aberrations corrected up to 30 mrad.

**Atomic force microscopy (AFM)**

Topographic characterization was performed using a Nanosurf Core AFM in tapping mode.

**Raman and Photoluminescence Spectroscopy**

The Raman and PL spectroscopy measurements were conducted using a confocal micro-Raman (PL) spectrometer (Horiba, LabRAM HR Evolution). The as-grown samples were excited using an excitation laser of 532 nm and collected the scattered radiation using x100 objective (0.9 NA). The scattered radiation was analyzed and detected using a grating of 1800 gr.mm$^{-1}$ and, thermoelectrically cooled CCD detector, respectively. The Raman and PL spectral imaging were performed on a software controlled XYZ motorized stage with a step resolution of ~100 nm. The laser power, exposure and collection timings were optimized to record the Raman and PL imaging. The incident laser power densities were calculated assuming the beam spot size (diameter) d= (1.22 λ / NA), where λ is the wavelength, and NA is is numerical aperture of the objective. The temperature dependent

PL measurements were performed in a cryostage (Linkam) cooled by Liq. $N_2$ under the vacuum environment and by using an objective of x50 (0.5 NA). Prior to the measurements, the samples were heated to 100 °c under vacuum for 30 min to remove the any adsorbed humidity and gas molecules.

**Time-resolved photoluminescence (TRPL) Spectroscopy measurements**

TRPL Spectroscopy measurements were performed using a home-built setup. The spectra were collected from the pristine samples on sapphire and the samples transferred on to quartz substrate. A pulsed laser of wavelength 405 nm (EPL-405nm, 55 ps pulse width) was used to excite the samples mounted on an inverted microscope (Zeiss). An oil-immersion objective lens (63×; NA: 1.4) was used to focus the incident laser and collect the PL signals in a confocal configuration and filtered by passing through the 488 nm dichroic mirror and 595 nm long pass filter. Prior to the TRPL, the point of interest was chosen by imaging onto a CCD (Andor, iXon Ultra). PL lifetimes were collected using a single photon avalanche photo diode (SPAD) (ID100, ID Quantique) coupled to a time-correlated single photon counter (TCSPC) module (HydraHarp 400, PicoQuant). The power dependent TRPL measurements were performed by varying the laser power using a set of ND filters and measuring the power by a powermeter (Nova II, Ophir).

**Time-of-flight secondary ion mass spectroscopy (TOF-SIMS) measurements**

Time-of-flight secondary ion mass spectroscopy (PHI 2100 TRIFT II) measurements were performed using $Ga^+$ ion beam. Surface mass spectra and depth profiles were collected. The depth profiling was performed in a dual-beam mode. A 15KeV $Ga^+$ (current of 600pA) ion beam was used as analytical beam, and a 500 eV $Cs^+$ (current of 50nA) ion beam was used for sputtering. Analyzed area was 50 x 50 $\mu m^2$, and sputtered area was 400 x 400 $\mu m^2$. The analysis was performed in the vacuum of 5 x $10^{-10}$ Torr, and in negative ions mode.

**References**

1. Lee, Y.-H.; Zhang, X.-Q.; Zhang, W.; Chang, M.-T.; Lin, C.-T.; Chang, K.-D.; Yu, Y.-C.; Wang, J. T.-W.; Chang, C.-S.; Li, L.-J.; Lin, T.-W., Synthesis of Large-Area MoS2 Atomic Layers with Chemical Vapor Deposition. *Adv. Mater.* **2012,** *24* (17), 2320-2325.

2. Hod, O.; Urbakh, M.; Naveh, D.; Bar-Sadan, M.; Ismach, A., Flatlands in the Holy Land: The Evolution of Layered Materials Research in Israel. *Adv. Mater.* **2018,** *0* (0), 1706581.
3. Eichfeld, S. M.; Hossain, L.; Lin, Y. C.; Piasecki, A. F.; Kupp, B.; Birdwell, A. G.; Burke, R. A.; Lu, N.; Peng, X.; Li, J.; Azcatl, A.; McDonnell, S.; Wallace, R. M.; Kim, M. J.; Mayer, T. S.; Redwing, J. M.; Robinson, J. A., Highly Scalable, Atomically Thin WSe2 Grown via Metal-Organic Chemical Vapor Deposition. *Acs Nano* **2015,** *9* (2), 2080-2087.
4. Yang, P.; Yang, A. G.; Chen, L. X.; Chen, J.; Zhang, Y. W.; Wang, H. M.; Hu, L. G.; Zhang, R. J.; Liu, R.; Qu, X. P.; Qiu, Z. J.; Cong, C. X., Influence of Seeding Promoters on the Pproperties of CVD Grown Monolayer Molybdenum Disulfide. *Nano Res.* **2019,** *12* (4), 823-827.
5. Teich, J.; Dvir, R.; Henning, A.; Hamo, E. R.; Moody, M. J.; Jurca, T.; Cohen, H.; Marks, T. J.; Rosen, B. A.; Lauhon, L. J.; Ismach, A., Light and Complex 3D MoS2/Graphene Heterostructures as Efficient Catalysts for the Hydrogen Evolution Reaction. *Nanoscale* **2020,** *12* (4), 2715-2725.
6. Liu, Y.; Ghosh, R.; Wu, D.; Ismach, A.; Ruoff, R.; Lai, K., Mesoscale Imperfections in MoS2 Atomic Layers Grown by a Vapor Transport Technique. *Nano Lett.* **2014,** *14* (8), 4682-4686.
7. Ling, X.; Lee, Y. H.; Lin, Y. X.; Fang, W. J.; Yu, L. L.; Dresselhaus, M. S.; Kong, J., Role of the Seeding Promoter in MoS2 Growth by Chemical Vapor Deposition. *Nano Lett.* **2014,** *14* (2), 464-472.
8. Patsha, A.; Sheff, V.; Ismach, A., Seeded-Growth of WS2 Atomic Layers: The Effect on Chemical and Optical Properties. *Nanoscale* **2019,** *11* (46), 22493-22503.
9. Yun, S. J.; Chae, S. H.; Kim, H.; Park, J. C.; Park, J.-H.; Han, G. H.; Lee, J. S.; Kim, S. M.; Oh, H. M.; Seok, J.; Jeong, M. S.; Kim, K. K.; Lee, Y. H., Synthesis of Centimeter-Scale Monolayer Tungsten Disulfide Film on Gold Foils. *ACS Nano* **2015,** *9* (5), 5510-5519.
10. Radovsky, G.; Shalev, T.; Ismach, A., Tuning the morphology and chemical composition of MoS2 nanostructures. *J. Mater. Sci.* **2019,** *54* (10), 7768-7779.
11. Jones, A. C.; Hitchman, M. L., *Chemical Vapour Deposition : Precursors, Processes and Applications*. Royal Society of Chemistry: Cambridge, UK, 2009.
12. Yu, H.; Liao, M.; Zhao, W.; Liu, G.; Zhou, X. J.; Wei, Z.; Xu, X.; Liu, K.; Hu, Z.; Deng, K.; Zhou, S.; Shi, J.-A.; Gu, L.; Shen, C.; Zhang, T.; Du, L.; Xie, L.; Zhu, J.; Chen, W.; Yang, R. *et al.*, Wafer-Scale Growth and Transfer of Highly-Oriented Monolayer MoS2 Continuous Films. *ACS Nano* **2017,** *11* (12), 12001-12007.
13. Mohapatra, P. K.; Ranganathan, K.; Ismach, A., Selective Area Growth and Transfer of High Optical Quality MoS2 Layers. *Adv. Mater. Interfaces* **2020**, 2001549.
14. Chiu, H. T.; Chuang, S. H., Tungsten Nitride Thin-Films Prepared by MOCVD *J. Mater. Res.* **1993,** *8* (6), 1353-1360.
15. Hofmann, W. K., Thin-Films of Molybdenum and Tungsten Disulfides by Metal Organic-Chemical Vapor-Deposition *J. Mater. Sci.* **1988,** *23* (11), 3981-3986.
16. Kang, K.; Xie, S.; Huang, L.; Han, Y.; Huang, P. Y.; Mak, K. F.; Kim, C.-J.; Muller, D.; Park, J., High-Mobility Three-Atom-Thick Semiconducting Films with Wafer-Scale Homogeneity. *Nature* **2015,** *520* (7549), 656-660.
17. Kalanyan, B.; Kimes, W. A.; Beams, R.; Stranick, S. J.; Garratt, E.; Kalish, I.; Davydov, A. V.; Kanjolia, R. K.; Maslar, J. E., Rapid Wafer-Scale Growth of

Polycrystalline 2H-MoS2 by Pulsed Metal–Organic Chemical Vapor Deposition. *Chem. Mater.* **2017,** *29* (15), 6279-6288.
18. Lee, D. H.; Sim, Y.; Wang, J.; Kwon, S. Y., Metal-Organic Chemical Vapor Deposition of 2D van der Waals Materials-The Challenges and the Extensive Future Opportunities. *Apl Materials* **2020,** *8* (3).
19. Cwik, S.; Mitoraj, D.; Reyes, O. M.; Rogalla, D.; Peeters, D.; Kim, J.; Schutz, H. M.; Bock, C.; Beranek, R.; Devi, A., Direct Growth of MoS2 and WS2 Layers by Metal Organic Chemical Vapor Deposition. *Adv. Mater. Interfaces* **2018,** *5* (16).
20. Andrzejewski, D.; Marx, M.; Grundmann, A.; Pfingsten, O.; Kalisch, H.; Vescan, A.; Heuken, M.; Kummell, T.; Bacher, G., Improved Luminescence Properties of MoS2 Monolayers Grown via MOCVD: Role of Pretreatment and Growth Parameters. *Nanotechnology* **2018,** *29* (29).
21. Marx, M.; Grundmann, A.; Lin, Y. R.; Andrzejewski, D.; Kümmell, T.; Bacher, G.; Heuken, M.; Kalisch, H.; Vescan, A., Metalorganic Vapor-Phase Epitaxy Growth Parameters for Two-Dimensional MoS2. *J. Electron. Mater.* **2018,** *47* (2), 910-916.
22. Choudhury, T. H.; Simchi, H.; Boichot, R.; Chubarov, M.; Mohney, S. E.; Redwing, J. M., Chalcogen Precursor Effect on Cold-Wall Gas-Source Chemical Vapor Deposition Growth of WS2. *Crystal Growth & Design* **2018,** *18* (8), 4357-4364.
23. Seol, M.; Lee, M.-H.; Kim, H.; Shin, K. W.; Cho, Y.; Jeon, I.; Jeong, M.; Lee, H.-I.; Park, J.; Shin, H.-J., High-Throughput Growth of Wafer-Scale Monolayer Transition Metal Dichalcogenide via Vertical Ostwald Ripening. *Adv. Mater. n/a* (n/a), 2003542.
24. Andrzejewski, D.; Myja, H.; Heuken, M.; Grundmann, A.; Kalisch, H.; Vescan, A.; Kümmell, T.; Bacher, G., Scalable Large-Area p–i–n Light-Emitting Diodes Based on WS2 Monolayers Grown via MOCVD. *ACS Photonics* **2019,** *6* (8), 1832-1839.
25. Xie, S. E.; Tu, L. J.; Han, Y. M.; Huang, L. J.; Kang, K.; Lao, K. U.; Poddar, P.; Park, C.; Muller, D. A.; DiStasio, R. A.; Park, J., Coherent, Atomically Thin Transition-Metal Dichalcogenide Superlattices with Engineered Strain. *Science* **2018,** *359* (6380), 1131-1135.
26. Choi, W.; Choudhary, N.; Han, G. H.; Park, J.; Akinwande, D.; Lee, Y. H., Recent Development of Two-Dimensional Transition Metal Dichalcogenides and their Applications. *Mater. Today* **2017,** *20* (3), 116-130.
27. Amani, M.; Burke, R. A.; Ji, X.; Zhao, P.; Lien, D.-H.; Taheri, P.; Ahn, G. H.; Kirya, D.; Ager, J. W.; Yablonovitch, E.; Kong, J.; Dubey, M.; Javey, A., High Luminescence Efficiency in MoS2 Grown by Chemical Vapor Deposition. *ACS Nano* **2016,** *10* (7), 6535-6541.
28. Amani, M.; Taheri, P.; Addou, R.; Ahn, G. H.; Kiriya, D.; Lien, D.-H.; Ager, J. W.; Wallace, R. M.; Javey, A., Recombination Kinetics and Effects of Superacid Treatment in Sulfur- and Selenium-Based Transition Metal Dichalcogenides. *Nano Lett.* **2016,** *16* (4), 2786-2791.
29. Tanoh, A. O. A.; Alexander-Webber, J.; Xiao, J.; Delport, G.; Williams, C. A.; Bretscher, H.; Gauriot, N.; Allardice, J.; Pandya, R.; Fan, Y.; Li, Z. J.; Vignolini, S.; Stranks, S. D.; Hofmann, S.; Rao, A., Enhancing Photoluminescence and Mobilities in WS2 Monolayers with Oleic Acid Ligands. *Nano Lett.* **2019,** *19* (9), 6299-6307.
30. Li, S. S.; Wang, S. F.; Tang, D. M.; Zhao, W. J.; Xu, H. L.; Chu, L. Q.; Bando, Y.; Golberg, D.; Eda, G., Halide-Assisted Atmospheric Pressure Growth of Large WSe2 and WS2 Monolayer Crystals. *Appl. Mater. Today* **2015,** *1* (1), 60-66.

31.	Kim, H.; Ovchinnikov, D.; Deiana, D.; Unuchek, D.; Kis, A., Suppressing Nucleation in Metal-Organic Chemical Vapor Deposition of MoS2 Monolayers by Alkali Metal Halides. *Nano Lett.* **2017,** *17* (8), 5056-5063.
32.	Zhang, K. H.; Bersch, B. M.; Zhang, F.; Briggs, N. C.; Subramanian, S.; Xu, K.; Chubarov, M.; Wang, K.; Lerach, J. O.; Redwing, J. M.; Fullerton-Shirey, S. K.; Terrones, M.; Robinson, J. A., Considerations for Utilizing Sodium Chloride in Epitaxial Molybdenum Disulfide. *ACS Appl. Mater. Interfaces* **2018,** *10* (47), 40831-40837.
33.	Utama, M. I. B.; Lu, X.; Yuan, Y. W.; Xiong, Q. H., Detrimental Influence of Catalyst Seeding on the Device Properties of CVD-grown 2D layered Materials: A Case Study on MoSe2. *Appl. Phys. Lett.* **2014,** *105* (25).
34.	Han, G. H.; Kybert, N. J.; Naylor, C. H.; Lee, B. S.; Ping, J.; Park, J. H.; Kang, J.; Lee, S. Y.; Lee, Y. H.; Agarwal, R.; Johnson, A. T. C., Seeded Ggrowth of Highly Crystalline Molybdenum Disulphide Monolayers at Controlled Locations. *Nat. Commun* **2015,** *6*.
35.	Koma, A., Vanderwaals epitaxy - A New Epitaxial method for Highly Lattice Mismatched System *Thin Solid Films* **1992,** *216* (1), 72-76.
36.	Zhang, X.; Choudhury, T. H.; Chubarov, M.; Xiang, Y.; Jariwala, B.; Zhang, F.; Alem, N.; Wang, G.-C.; Robinson, J. A.; Redwing, J. M., Diffusion-Controlled Epitaxy of Large Area Coalesced WSe2 Monolayers on Sapphire. *Nano Lett.* **2018,** *18* (2), 1049-1056.
37.	Mattinen, M.; King, P. J.; Popov, G.; Hamalainen, J.; Heikkila, M. J.; Leskela, M.; Ritala, M., Van der Waals Epitaxy of Continuous Thin Films of 2D Materials Using Atomic Layer Deposition in Low Temperature and Low Vacuum Conditions. *2d Materials* **2020,** *7* (1).
38.	Zhang, K. H.; Jariwala, B.; Li, J.; Briggs, N. C.; Wang, B. M.; Ruzmetov, D.; Burke, R. A.; Lerach, J. O.; Ivanov, T. G.; Haque, M.; Feenstra, R. M.; Robinson, J. A., Large Scale 2D/3D hybrids Based on Gallium Nitride and Transition metal Dichalcogenides. *Nanoscale* **2018,** *10* (1), 336-341.
39.	Fu, W.; Qiao, J. S.; Zhao, X. X.; Chen, Y.; Fu, D. Y.; Yu, W.; Leng, K.; Song, P.; Chen, Z.; Yu, T.; Pennycook, S. J.; Quek, S. Y.; Loh, K. P., Room Temperature Commensurate Charge Density Wave on Epitaxially Grown Bilayer 2H-Tantalum Sulfide on Hexagonal Boron Nitride. *Acs Nano* **2020,** *14* (4), 3917-3926.
40.	Liu, X. L.; Balla, I.; Bergeron, H.; Campbell, G. P.; Bedzyk, M. J.; Hersam, M. C., Rotationally Commensurate Growth of MoS2 on Epitaxial Graphene. *Acs Nano* **2016,** *10* (1), 1067-1075.
41.	Zhang, X.; Meng, F.; Christianson, J. R.; Arroyo-Torres, C.; Lukowski, M. A.; Liang, D.; Schmidt, J. R.; Jin, S., Vertical Heterostructures of Layered Metal Chalcogenides by van der Waals Epitaxy. *Nano Lett.* **2014,** *14* (6), 3047-3054.
42.	Mohapatra, P. K.; Ranganathan, K.; Dezanashvili, L.; Houben, L.; Ismach, A., Epitaxial Growth of In2Se3 on Monolayer Transition Metal Dichalcogenide Single Crystals for High Performance Photodetectors. *Appl. Mater. Today* **2020,** *20*, 100734.
43.	Dhar, S.; Kranthi Kumar, V.; Choudhury, T. H.; Shivashankar, S. A.; Raghavan, S., Chemical Vapor Deposition of MoS2 Layers from Mo–S–C–O–H System: Thermodynamic Modeling and Validation. *PCCP* **2016,** *18* (22), 14918-14926.
44.	Kumar, V. K.; Dhar, S.; Choudhury, T. H.; Shivashankar, S. A.; Raghavan, S., A Predictive Approach to CVD of Crystalline Layers of TMDs: The Case of MoS2. *Nanoscale* **2015,** *7* (17), 7802-7810.

45. Momeni, K.; Ji, Y.; Wang, Y.; Paul, S.; Neshani, S.; Yilmaz, D. E.; Shin, Y. K.; Zhang, D.; Jiang, J.-W.; Park, H. S.; Sinnott, S.; van Duin, A.; Crespi, V.; Chen, L.-Q., Multiscale Computational Understanding and Growth of 2D Materials: a Review. *npj Computational Materials* **2020,** *6* (1), 22.
46. Shang, S.-L.; Lindwall, G.; Wang, Y.; Redwing, J. M.; Anderson, T.; Liu, Z.-K., Lateral *Versus* Vertical Growth of Two-Dimensional Layered Transition-Metal Dichalcogenides: Thermodynamic Insight into MoS2. *Nano Lett.* **2016,** *16* (9), 5742-5750.
47. Choudhury, T. H.; Zhang, X. T.; Al Balushi, Z. Y.; Chubarov, M.; Redwing, J. M., Epitaxial Growth of Two-Dimensional Layered Transition Metal Dichalcogenides. In *Annual Review of Materials Research, Vol 50, 2020*, Clarke, D. R., Ed. 2020; Vol. 50, pp 155-177.
48. Zhang, X. T.; Balushi, Z. Y.; Zhang, F.; Choudhury, T. H.; Eichfeld, S. M.; Alem, N.; Jackson, T. N.; Robinson, J. A.; Redwing, J. M., Influence of Carbon in Metalorganic Chemical Vapor Deposition of Few-Layer WSe2 Thin Films. *J. Electron. Mater.* **2016,** *45* (12), 6273-6279.
49. Eichfeld, S. M.; Colon, V. O.; Nie, Y. F.; Cho, K.; Robinson, J. A., Controlling Nucleation of Monolayer WSe2 During Metal-Organic Chemical Vapor Deposition Growth. *2d Materials* **2016,** *3* (2).
50. Choi, S. H.; Stephen, B.; Park, J. H.; Lee, J. S.; Kim, S. M.; Yang, W.; Kim, K. K., Water-Assisted Synthesis of Molybdenum Disulfide Film with Single Organic Liquid Precursor. *Sci. Rep.* **2017,** *7*.
51. Kastl, C.; Chen, C. T.; Kuykendall, T.; Shevitski, B.; Darlington, T. P.; Borys, N. J.; Krayev, A.; Schuck, P. J.; Aloni, S.; Schwartzberg, A. M., The Important Role of Water in Growth of Monolayer Transition Mmetal Dichalcogenides. *2d Materials* **2017,** *4* (2).
52. Berkdemir, A.; Gutiérrez, H. R.; Botello-Méndez, A. R.; Perea-López, N.; Elías, A. L.; Chia, C.-I.; Wang, B.; Crespi, V. H.; López-Urías, F.; Charlier, J.-C.; Terrones, H.; Terrones, M., Identification of Individual and Few Layers of WS2 Using Raman Spectroscopy. *Sci. Rep.* **2013,** *3* (1), 1755.
53. Krisyuk, V. V.; Koretskaya, T. P.; Turgambaeva, A. E.; Trubin, S. V.; Korolkov, I. V.; Debieu, O.; Duguet, T.; Igumenov, I. K.; Vahlas, C., Thermal Decomposition of Tungsten Hexacarbonyl: CVD of W-Containing Films Under Pd Codeposition and VUV Assistance. *Phys. Status Solidi C* **2015,** *12* (7), 1047-1052.
54. Cannon, P.; Norton, F. J., Reaction Between Molybdenum Disulphide and Water. *Nature* **1964,** *203* (4946), 750-751.
55. Millner, T.; Neugebauer, J., Volatility of the Oxides of Tungsten and Molybdenum in the Presence of Water Vapour. *Nature* **1949,** *163* (4146), 601-602.
56. Gurarslan, A.; Yu, Y.; Su, L.; Yu, Y.; Suarez, F.; Yao, S.; Zhu, Y.; Ozturk, M.; Zhang, Y.; Cao, L., Surface-Energy-Assisted Perfect Transfer of Centimeter-Scale Monolayer and Few-Layer MoS2 Films onto Arbitrary Substrates. *ACS Nano* **2014,** *8* (11), 11522-11528.
57. Lenz, M.; Gruehn, R., Developments in Measuring and Calculating Chemical Vapor Transport Phenomena Demonstrated on Cr, Mo, W, and Their Compounds. *Chem. Rev.* **1997,** *97* (8), 2967-2994.
58. Glemser, O.; Haeseler, R. V., Gasförmige Hydroxide. IV. Über gasförmige Hydroxide des Molybdäns und Wolframs. *Z. Anorg. Allg. Chem.* **1962,** *316* (3-4), 168-181.

59. You, Y.; Zhang, X.-X.; Berkelbach, T. C.; Hybertsen, M. S.; Reichman, D. R.; Heinz, T. F., Observation of Biexcitons in Monolayer WSe2. *Nat. Phys.* **2015,** *11* (6), 477-481.
60. Tongay, S.; Suh, J.; Ataca, C.; Fan, W.; Luce, A.; Kang, J. S.; Liu, J.; Ko, C.; Raghunathanan, R.; Zhou, J.; Ogletree, F.; Li, J.; Grossman, J. C.; Wu, J., Defects Activated Photoluminescence in Two-Dimensional Semiconductors: Iinterplay Between Bound, Charged and Free Excitons. *Sci. Rep.* **2013,** *3* (1), 2657.
61. Shang, J.; Shen, X.; Cong, C.; Peimyoo, N.; Cao, B.; Eginligil, M.; Yu, T., Observation of Excitonic Fine Structure in a 2D Transition-Metal Dichalcogenide Semiconductor. *ACS Nano* **2015,** *9* (1), 647-655.
62. Mak, K. F.; He, K.; Lee, C.; Lee, G. H.; Hone, J.; Heinz, T. F.; Shan, J., Tightly Bound Trions in Monolayer MoS2. *Nat. Mater.* **2013,** *12* (3), 207-211.
63. Bellus, M. Z.; Ceballos, F.; Chiu, H.-Y.; Zhao, H., Tightly Bound Trions in Transition Metal Dichalcogenide Heterostructures. *ACS Nano* **2015,** *9* (6), 6459-6464.
64. Carozo, V.; Wang, Y.; Fujisawa, K.; Carvalho, B. R.; McCreary, A.; Feng, S.; Lin, Z.; Zhou, C.; Perea-López, N.; Elías, A. L.; Kabius, B.; Crespi, V. H.; Terrones, M., Optical identification of sulfur vacancies: Bound Excitons at the Edges of Monolayer Tungsten Disulfide. *Sci. Adv.* **2017,** *3* (4), e1602813.
65. He, Z.; Wang, X.; Xu, W.; Zhou, Y.; Sheng, Y.; Rong, Y.; Smith, J. M.; Warner, J. H., Revealing Defect-State Photoluminescence in Monolayer WS2 by Cryogenic Laser Processing. *ACS Nano* **2016,** *10* (6), 5847-5855.
66. Chow, P. K.; Jacobs-Gedrim, R. B.; Gao, J.; Lu, T.-M.; Yu, B.; Terrones, H.; Koratkar, N., Defect-Induced Photoluminescence in Monolayer Semiconducting Transition Metal Dichalcogenides. *ACS Nano* **2015,** *9* (2), 1520-1527.
67. Amani, M.; Lien, D.-H.; Kiriya, D.; Xiao, J.; Azcatl, A.; Noh, J.; Madhvapathy, S. R.; Addou, R.; KC, S.; Dubey, M.; Cho, K.; Wallace, R. M.; Lee, S.-C.; He, J.-H.; Ager, J. W.; Zhang, X.; Yablonovitch, E.; Javey, A., Near-Unity Photoluminescence Quantum Yield in MoS2. *Science* **2015,** *350* (6264), 1065-1068.
68. Mignuzzi, S.; Pollard, A. J.; Bonini, N.; Brennan, B.; Gilmore, I. S.; Pimenta, M. A.; Richards, D.; Roy, D., Effect of disorder on Raman Scattering of Single-Layer MoS2. *Physical Review B* **2015,** *91* (19), 195411.
69. McCreary, A.; Berkdemir, A.; Wang, J.; Nguyen, M. A.; Elías, A. L.; Perea-López, N.; Fujisawa, K.; Kabius, B.; Carozo, V.; Cullen, D. A.; Mallouk, T. E.; Zhu, J.; Terrones, M., Distinct Photoluminescence and Raman Spectroscopy Signatures for Identifying Highly Crystalline WS2 Monolayers Produced by Different Growth Methods. *J. Mater. Res.* **2016,** *31* (7), 931-944.
70. Sun, D.; Rao, Y.; Reider, G. A.; Chen, G.; You, Y.; Brézin, L.; Harutyunyan, A. R.; Heinz, T. F., Observation of Rapid Exciton–Exciton Annihilation in Monolayer Molybdenum Disulfide. *Nano Lett.* **2014,** *14* (10), 5625-5629.
71. Lee, Y.; Ghimire, G.; Roy, S.; Kim, Y.; Seo, C.; Sood, A. K.; Jang, J. I.; Kim, J., Impeding Exciton–Exciton Annihilation in Monolayer WS2 by Laser Irradiation. *ACS Photonics* **2018,** *5* (7), 2904-2911.
72. Yuan, L.; Huang, L., Exciton Dynamics and Annihilation in WS2 2D Semiconductors. *Nanoscale* **2015,** *7* (16), 7402-7408.
73. Zhang, X. X.; You, Y. M.; Zhao, S. Y. F.; Heinz, T. F., Experimental Evidence for Dark Excitons in Monolayer WSe2. *Phys. Rev. Lett.* **2015,** *115* (25).